\documentclass[a4paper, 12pt]{report}

\usepackage[english]{babel}
\usepackage[noheader]{packages/sleek}
\usepackage{packages/sleek-title}
\usepackage{packages/sleek-theorems}
\usepackage{packages/sleek-listings}
\usepackage{longtable} 
\usepackage{tabularx}
\usepackage{ragged2e}
\usepackage{booktabs}
\usepackage{soul}
\usepackage{comment}
\sethlcolor{lightgray}

\institute{TCG CREST}
\faculty{Institute of Advancing Intelligence}
\title{DeepOps \& SLURM: Your GPU Cluster Guide}
\subtitle{A guidance to build your own GPU cluster for Deep Learning using SLURM \& NVIDIA Deep-ops}
\author{\textit{Author}\\Arindam Majee \\ {\footnotesize Institute of Advancing Intelligence, TCG CREST \\ Email- majeearindam06072002@gmail.com}}
\supervisor{\textit{Supervisor}\\Prof. Swagatam Das \\{\footnotesize Indian Statistical Institute}}
\date{\today}

\lstdefinestyle{latex}{
    language=TeX,
    style=default,
    commentstyle=\ForestGreen,
    keywordstyle=\TrueBlue,
    stringstyle=\VeronicaPurple,
    emphstyle=\TrueBlue,
    emph={LaTeX, usepackage, textit, textbf, textsc}
}

\FrameTBStyle{latex}

\begin{document}
    \maketitle
    \romantableofcontents

    \chapter{Introduction}

    In the ever-evolving landscape of deep learning, unlocking the potential of cutting-edge models demands computational resources that surpass the capabilities of individual machines. Enter the NVIDIA DeepOps Slurm cluster – a meticulously orchestrated symphony of high-performance nodes, each equipped with powerful GPUs and meticulously managed by the efficient Slurm resource allocation system. This guide serves as your comprehensive roadmap, empowering you to harness the immense parallel processing capabilities of this cluster and propel your deep learning endeavors to new heights.

    Whether you're a seasoned deep learning practitioner seeking to optimize performance or a newcomer eager to unlock the power of parallel processing, this guide caters to your needs. We'll delve into the intricacies of the cluster's hardware architecture, exploring the capabilities of its GPUs and the underlying network fabric. You'll master the art of leveraging DeepOps containers for efficient and reproducible workflows, fine-tune resource configurations for optimal performance, and confidently submit jobs to unleash the full potential of parallel processing.
    
    No prior experience with clusters or deep learning is necessary. We'll guide you through every step with clear, concise explanations and practical examples. By the end of this journey, you'll be equipped with the knowledge and skills to:

    \begin{itemize}
        \item \textbf{Demystify the cluster architecture}: Understand the hardware components, network topology, and communication protocols that power your cluster.
        \item \textbf{Harness DeepOps containers}: Leverage their benefits for streamlined workflow management and environment consistency.
        \item \textbf{Optimize resource allocation}: Fine-tune configurations to match your specific deep learning workloads and maximize performance.
        \item \text: Master the art of crafting effective job scripts and utilizing Slurm's capabilities for efficient job execution.
    \end{itemize}
    
    Embark on this transformative journey with us, and unlock the immense potential of your DeepOps Slurm cluster. Let's empower your deep learning ambitions and push the boundaries of AI innovation, together.

    \chapter{Preliminary Concepts}
    Welcome to the exciting world of deep learning with your NVIDIA DeepOps Slurm cluster! Before we dive into the intricate workings of your powerful tool, let's establish a solid foundation in some key concepts. This chapter will equip you with the essential knowledge to understand and effectively utilize your cluster, even if you're new to deep learning or cluster computing.

    \section{What is Deep Learning?}
    Deep learning is a subset of artificial intelligence (AI) that utilizes artificial neural networks inspired by the structure and function of the human brain. These networks learn and improve by processing large amounts of data, enabling them to perform tasks such as image recognition, natural language processing, and even generating creative content.
    
    \section{Deep Learning and the Limits of Single Machines}
    Deep learning has transformed our world, from recognizing faces in photos to translating languages on-the-fly, and even diagnosing medical conditions with remarkable accuracy. But as these models become increasingly complex, they're facing a critical roadblock: the limitations of single machines. This section explores the growing tension between the ambitious goals of deep learning and the physical constraints of our computational resources.

    \subsection{The Power and Hunger of Deep Learning}
    Deep learning models excel at mimicking the human brain's ability to learn from vast amounts of data. By layering artificial neurons, these models extract intricate patterns and relationships, enabling them to perform tasks previously considered impossible for machines. However, this power comes at a cost. Deep learning models are notoriously resource-intensive, demanding massive amounts of computing power and memory to train and operate effectively.

    \subsection{The Bottleneck of Single Machines}
    Traditional computers, while constantly improving, have inherent limitations. Processors have reached a point of diminishing returns in terms of clock speed, and single-core performance gains have stagnated. Meanwhile, the memory requirements of deep learning models are skyrocketing, often exceeding the capacity of even the most powerful workstations.
    
    This growing mismatch between model complexity and hardware capabilities presents a significant challenge for the future of deep learning. Several consequences arise:
    \begin{itemize}
        \item Slower Progress: Training increasingly complex models on single machines becomes prohibitively slow, hindering research and development.
        \item Accessibility Bottleneck: The high computational cost creates a barrier to entry, limiting access to cutting-edge deep learning for smaller research groups and startups.
        \item Environmental Concerns: The immense power consumption of training deep learning models raises environmental concerns, pushing for more efficient solutions.
    \end{itemize}
    
    \subsection{Beyond the Walls of Single Machines}
    To overcome these limitations, researchers and developers are exploring various avenues:
    \begin{itemize}
        \item Distributed Computing: Harnessing the collective power of multiple computers, such as clusters or cloud platforms, allows for parallelization of training and inference, significantly speeding up the process.
        \item Model Compression: Techniques like pruning and quantization can reduce the size and complexity of deep learning models, making them run more efficiently on single machines.
        \item Hardware Innovation: New hardware architectures, such as neuromorphic chips and quantum computers, hold promise for significantly boosting computational power and efficiency, specifically tailored for deep learning tasks.
    \end{itemize}

    The future of deep learning lies not just in pushing the boundaries of model complexity but also in finding innovative ways to bridge the gap between these models and the hardware that runs them. Collaboration between researchers, developers, and hardware manufacturers is crucial to ensure continued progress in this field. By embracing distributed computing, exploring novel hardware, and developing more efficient models, we can ensure that deep learning continues its transformative journey, even as we reach the limits of single machines.

    \section{Single computer vs. cluster for deep learning}
    \begin{table}[htbp]
    \centering
    \begin{tabularx}{\textwidth}{|>{\RaggedRight\arraybackslash}X|>{\RaggedRight\arraybackslash}X|>{\RaggedRight\arraybackslash}X|}
    \hline
    \textbf{Feature} & \textbf{Single Computer} & \textbf{Cluster} \\ \hline
    Computational Power & Limited CPU cores, single GPU & Scalable, parallel processing with multiple nodes \\ \hline
    Training Speed & Slow, sequential processing & Significantly faster, parallel processing \\ \hline
    Memory & Limited, bottlenecks with large models & Can handle large models, combined resources \\ \hline
    Cost & Inexpensive, good for small projects & Expensive, but cost-effective for complex tasks \\ \hline
    Flexibility & Limited configuration options & Highly customizable, optimized for specific tasks \\ \hline
    Data Management & Local storage, inconvenient for large datasets & Shared storage, efficient access and collaboration \\ \hline
    Complexity & Easy setup and management & Requires technical expertise for setup and management \\ \hline
    Scalability & Difficult to scale beyond single machine & Easily scalable by adding nodes \\ \hline
    Suitable for & Individual learning, small projects & Large projects, complex models, research, real-world applications \\ \hline
    \end{tabularx}
    \caption{Single Computer vs. Cluster for Deep Learning Training: A Comparison}
    \label{tab:my_label}
    \end{table}

    Training a deep learning model on a single machine is akin to this single-handed struggle. It can take days, weeks, or even months for the model to crunch through the vast amounts of data and refine its parameters. This lengthy training time hinders development, experimentation, and ultimately, progress in various fields like computer vision, natural language processing, and robotics.

    Here's why single machines fall short:
    \begin{itemize}
        \item Limited CPU Cores: Modern CPUs, while powerful, have a limited number of cores (typically 4-16). Deep learning algorithms, however, are inherently parallel, meaning they can break down tasks into smaller subtasks and process them simultaneously. With limited cores, only a few subtasks can be handled at once, creating a bottleneck in the training process.
        
        \item Memory Constraints: Deep learning models often require large amounts of memory to store data, intermediate calculations, and the model itself. Single machines, especially desktops, might not have enough memory to handle these demands, leading to frequent data swapping and further slowing down training.
        
        \item GPU Bottlenecks: While CPUs handle general-purpose computations, GPUs excel at the matrix operations and convolutions that are the bread and butter of deep learning. However, most single machines have only one or two GPUs, restricting the parallel processing power available for training.
    \end{itemize}
    
    These limitations of single machines highlight the need for a more efficient approach to deep learning training. This is where the mighty cluster comes in.

    \section{Unleashing the Power of Clusters: A Deep Dive into Parallelism}
    Imagine a grand orchestra, each instrument playing its part in perfect harmony, creating a masterpiece of sound. Now, imagine that each instrument represents a node in your DeepOps Slurm cluster, and the conductor is the masterful Slurm resource manager. This analogy captures the essence of parallelism, the secret weapon that unlocks the true power of your cluster for deep learning.

    Forget the limitations of single machines. Parallelism allows your cluster to break down complex tasks into smaller, manageable chunks, distributing them across its multiple nodes. This synchronized dance of computation accelerates training, empowers larger models, and ultimately, opens doors to groundbreaking AI achievements.
    
    \subsection{Parallelism for Deep Learning:}
    
    In deep learning, we break down complex tasks into smaller, independent subtasks. These subtasks can be processed simultaneously on different nodes, significantly reducing the overall training time. Let's explore the two main types of parallelism used in deep learning on clusters:
        \begin{itemize}
            \item \textbf{Data Parallelism}: This approach splits the training data into smaller batches and distributes them across different nodes. Each node trains its own copy of the model on its assigned batch, and the results are then averaged to update the global model. Imagine having multiple chefs cooking the same dish using different ingredients, then combining the best elements for the final masterpiece.
    
            \item \textbf{Model Parallelism}: This strategy divides the deep learning model itself into smaller parts and assigns them to different nodes. Each node trains its assigned portion of the model on the entire training data. This approach is particularly effective for handling massive models that wouldn't fit on a single node. Think of it as specialized chefs working on different sections of a giant cake, ensuring every layer is perfectly baked before assembling the final masterpiece.
        \end{itemize}
    
    \subsection{Benefits of Parallelism for Deep Learning:}
        \begin{itemize}
            \item Faster Training: By distributing the workload, parallelism significantly reduces training time. Imagine finishing your meal in minutes instead of hours!
            \item Large Model Training: Parallelism allows us to train massive models that wouldn't be possible on a single machine. This opens doors to more powerful and complex AI applications.
            \item Efficient Resource Utilization: Clusters enable efficient utilization of hardware resources like CPU cores and GPUs, maximizing their potential for deep learning tasks.
        \end{itemize}
    
    \subsection{Challenges and Considerations:}
        \begin{itemize}
            \item Communication Overhead: As nodes work in parallel, communicating and synchronizing their progress can introduce overhead. This needs to be optimized for efficient training.
            \item Data Management: Efficient access and distribution of training data across the cluster is crucial for optimal performance.
            \item Job Scheduling and Management: Effectively scheduling and managing jobs across multiple nodes is essential for maximizing cluster utilization and avoiding bottlenecks.
        \end{itemize}

    \subsection{Beyond Brute Force: Deep Learning-Specific Benefits of Clusters}
    While clusters undeniably offer raw computational power, their true value for deep learning lies in their ability to amplify the unique needs of this demanding field. Unlike generic workloads, deep learning algorithms thrive on a specific set of advantages that clusters deliver, pushing the boundaries of what's possible in AI development and deployment:
    \begin{itemize}
        \item Memory Liberation: Deep learning models, especially intricate ones with multiple layers and billions of parameters, often devour vast amounts of memory. Single machines, even powerful ones, can choke under such demands, hindering training and experimentation. Clusters pool the memory resources of multiple nodes, creating a vast, unified pool that seamlessly accommodates even the most memory-hungry models. This freedom from memory constraints allows researchers and developers to explore larger, more complex architectures with higher accuracy and performance.

        \item  Data Deluge Dominion: Deep learning thrives on data, and the bigger the dataset, the more sophisticated the model it can learn. However, handling massive datasets on single machines becomes a logistical nightmare, with limitations on storage and processing speed. Clusters, with their distributed storage capabilities and parallel processing power, conquer this challenge. They can efficiently store and distribute large datasets across nodes, allowing each node to access and process its assigned portion simultaneously. This significantly speeds up training times, enabling the analysis of vast datasets that would be impossible on a single machine, leading to breakthroughs in areas like image recognition, natural language processing, and more.

        \item GPU Symphony: Deep learning algorithms rely heavily on the parallel processing power of GPUs to efficiently perform matrix operations and convolutions. While single machines may have one or two GPUs, their limitations become evident in complex tasks. Clusters can house dozens, even hundreds, of GPUs working in unison, creating a powerful symphony of parallel processing that dramatically accelerates the training process. Imagine training a complex model in hours, not weeks, on a cluster, allowing for faster iteration and experimentation, ultimately leading to the development of more efficient and accurate models.

        \item Collaboration and Scalability: Deep learning is often a collaborative effort, with teams of researchers and engineers working together. Traditional methods of sharing data and models can be cumbersome and inefficient. Clusters facilitate collaboration by providing a shared platform where data, models, and results can be easily accessed and shared. This fosters open communication and knowledge exchange, leading to faster progress and innovation.

        \item Cluster Advantage: Clusters are inherently scalable. Adding new nodes seamlessly expands the computational power and data storage capabilities, allowing users to adapt their resources to the growing demands of their projects. This flexibility ensures that their cluster can evolve alongside their deep learning ambitions.
    \end{itemize}
    
    In conclusion, clusters go beyond simply offering brute computational power for deep learning. They provide a unique ecosystem that unlocks the field's full potential by addressing its specific memory, data, GPU, and collaboration needs. By leveraging these deep learning-specific advantages, clusters empower researchers and developers to push the boundaries of AI, leading to groundbreaking advancements in various fields, from medical imaging to autonomous vehicles. As deep learning continues to evolve, clusters will remain an essential tool, enabling us to tackle even more complex challenges and unlock the true potential of artificial intelligence. Dive into the upcoming chapters and discover how your DeepOps Slurm cluster can become your springboard to transformative AI achievements!

    \chapter{Cluster Architecture}
    \section{Hardware Components}
    \subsection{Compute Nodes}
        These are individual machines within the cluster responsible for performing computations.
        Nodes are equipped with processors (CPUs), memory (RAM), and in deep learning clusters, Graphics Processing Units (GPUs) are often crucial for accelerated model training.

    \subsection{GPUs for Deep Learning}
\begin{figure}
            \centering
            \includegraphics[scale=0.3]{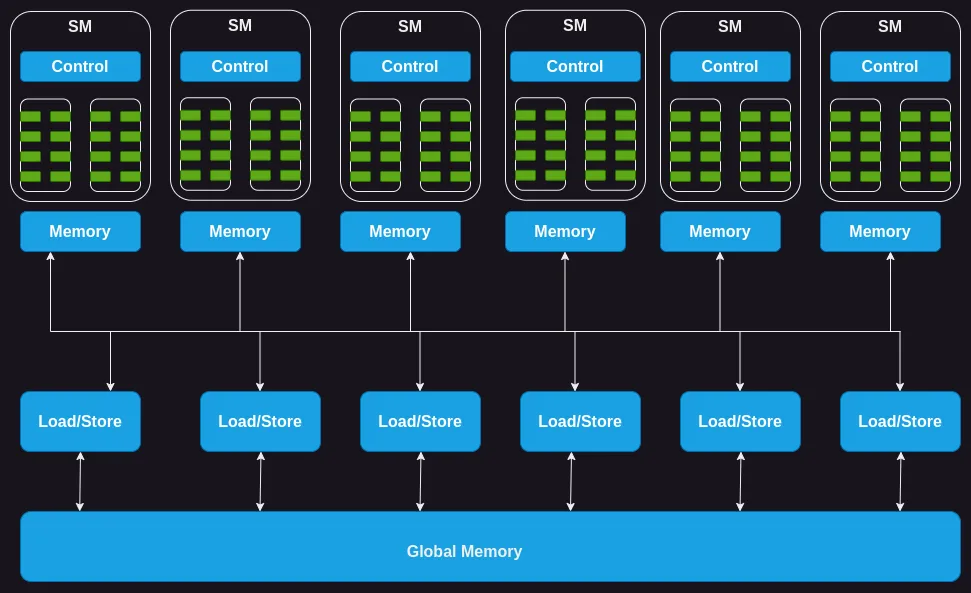}
            \caption{The GPU Compute Architecture}
            \label{fig:gpu_architecture}
        \end{figure}

        NVIDIA GPUs are widely used in deep learning clusters due to their parallel processing capabilities ideal for training neural networks. GPUs excel in handling matrix and vector operations, essential for deep learning computations.
        A GPU consists of an array of streaming multiprocessors (SM). Look at the Figure \ref{fig:gpu_architecture} for an overview of GPU compute architecture. Each of these SMs in turn consists of several streaming processors or cores or threads. For instance, the Nvidia H100 GPU has 132 SMs with 64 cores per SM, totalling a whopping 8448 cores.
        Each SM has a limited amount of on-chip memory, often referred to as shared memory or a scratchpad, which is shared among all the cores. Likewise, the control unit resources on the SM are shared by all the cores. Additionally, each SM is equipped with hardware-based thread schedulers for executing threads.
        
        Apart from these, each SM also has several functional units or other accelerated compute units such as tensor cores, or ray tracing units to serve specific compute demands of the workload that the GPU caters to.

    \subsection{Networking Infrastructure}
        The beating heart of your NVIDIA DeepOps Slurm cluster lies not in the individual nodes, but in the intricate network that seamlessly connects them. This network infrastructure serves as the vital conduit for data, the conductor orchestrating tasks across the entire system, and ultimately, the key to unlocking the true potential of your deep learning endeavors.
        \subsubsection{High-Speed Interconnects: Beyond Traditional Ethernet}
            Forget the limitations of standard Ethernet. Your DeepOps cluster demands a network built for speed and efficiency. Enter the world of high-performance interconnects like NVIDIA Mellanox InfiniBand or NVIDIA Spectrum Ethernet. These specialized networks offer a dramatic leap forward in both bandwidth and latency, boasting speeds of up to 200 Gb/s per link and drastically lower latencies compared to standard Ethernet. This translates to smooth and uninterrupted data flow, even when handling massive datasets and complex computations – essential for the demanding workloads of deep learning. If you don't have InfiniBand or Omni-Path supports (as they are costly), you may proceed with high-speed Ethernets.
            
        \subsubsection{Full-Mesh Topology: Redundancy and Resilience}
            While traditional full-mesh networks offer unparalleled redundancy and direct connectivity, it also offers lots of complications as the number of nodes increases. Hence, the DeepOps Slurm cluster often utilizes a different approach: the star topology. This configuration features a central switch (or router), acting as the hub, with each individual node directly connected to it.
            This star topology offers several advantages tailored to deep learning workloads:
            \begin{itemize}
                \item Scalability: Adding new nodes is as simple as plugging them into the central switch, making it easy to expand your cluster's computational power as your needs evolve.
                \item Cost-effectiveness: Compared to the dense cabling required in a full-mesh, the star topology uses less hardware, reducing initial setup costs.
                \item Managed Complexity: The central switch simplifies network management and troubleshooting, as all traffic flows through a single point.
                \item Efficient Communication: While not offering the complete connectivity of full-mesh, the star topology still provides efficient communication for many deep learning tasks. The central switch intelligently routes data, ensuring information reaches its destination quickly and reliably.
            \end{itemize}
            
        \subsubsection{Smart Switches: The Network's Traffic Controllers}
            High-performance fabric switches are the ideal choice for DeepOps clusters. They boast specialized features like:
            \begin{itemize}
                \item Low Latency: Designed to minimize delays, ensuring data travels quickly between nodes, crucial for real-time communication in deep learning tasks.
                \item High Throughput: Wide bandwidth allows for efficient transfer of massive datasets, essential for training complex models.
                \item Intelligent Routing: Employ sophisticated algorithms to optimize data paths and avoid congestion, maximizing network efficiency.
            \end{itemize}

            While fabric switches offer optimal performance, budget constraints may necessitate exploring alternatives. Routers can be used, but they often have higher latency and lower throughput compared to fabric switches. This might be acceptable for smaller clusters or less demanding workloads, but it can impact performance for larger clusters or complex tasks.  Creating a dedicated routing table for your cluster nodes can enhance security by isolating their network traffic from other systems and reduce communication overhead.

        \subsubsection{Deep Learning-Specific Optimizations: Going Beyond the Basics}
            But the DeepOps network doesn't stop at just high-performance hardware. It's further fine-tuned to cater to the specific needs of deep learning workloads:
            \begin{itemize}
                \item GPU Direct Communication: Imagine GPUs directly talking to each other, bypassing the CPU and slashing communication time. This is the magic of GPU Direct Communication, enabling efficient distributed training of deep learning models where data and gradients need to flow rapidly between nodes.
                \item RDMA (Remote Direct Memory Access): This technology allows GPUs to directly access memory on other nodes, eliminating unnecessary data copying and further boosting efficiency. Think of it as GPUs having their own personal storage lockers across the network, instantly accessing the data they need without waiting in line.
                \item Network Function Offloading: Specialized network adapters can take over tasks like packet processing and checksum calculations, freeing up valuable CPU resources for your core deep learning computations. It's like having dedicated assistants handle the network traffic, allowing your CPUs to focus on the critical thinking – training your models.
            \end{itemize}
            
            By understanding and appreciating the intricate workings of your DeepOps cluster's network infrastructure, you gain a deeper understanding of its capabilities and unlock its full potential. Remember, the network is not just a supporting player; it's the conductor of the deep learning symphony, orchestrating the seamless flow of information and paving the way for groundbreaking discoveries.

    \subsection{Storage Solutions}
        In the world of deep learning, data is king. But where does this king reside in cluster? The answer lies in your cluster's storage solutions, the foundation upon which your deep learning endeavors rest. Your storage needs are multifaceted. On one hand, you require high-performance for rapid data access during training and inference. This is where solid-state drives (SSDs) come in, offering significantly faster read/write speeds compared to traditional hard disk drives (HDDs). On the other hand, you need ample capacity to store massive datasets, model checkpoints, and training logs. This is where network-attached storage (NAS) excels, providing centralized storage accessible across the entire cluster.

        \subsubsection{Choosing the Right Storage Mix} 
        The optimal storage mix depends on your specific needs and budget:
        \begin{itemize}
            \item High-performance SSDs: Ideal for frequently accessed data like training datasets and active models. Consider NVMe-based SSDs for the ultimate speed boost.
            \item NAS: Great for storing large, less frequently accessed data archives and model checkpoints. Look for NAS solutions with high bandwidth and low latency for optimal performance.
            \item Distributed File Systems: With multiple nodes in your cluster, ensuring everyone has access to the data is crucial. Distributed file systems (DFS) make this possible, creating a single, unified view of your data across all nodes. This enables seamless collaboration and efficient data access for all your deep learning tasks. Popular choices for DeepOps clusters include:
            \begin{itemize}
                \item Lustre: High-performance DFS designed for large-scale HPC and deep learning workloads.
                \item GlusterFS: Open-source, scalable DFS offering good performance and flexibility.
                \item Ceph: Highly scalable and distributed object storage solution that can also be used as a DFS.
                \item BeeGFS: BeeGFS is a hardware-independent POSIX parallel file system (a.k.a Software-defined Parallel Storage), developed and optimized for HPC. 
            \end{itemize}
            \item Hybrid storage: Combine SSDs for speed-critical data with NAS and DFS for archival storage, offering a cost-effective balance.
        \end{itemize}

By carefully considering your storage needs, choosing the right mix of solutions, and implementing effective data management practices, you can build a robust foundation for your deep learning journey. Remember, a well-managed storage system empowers your DeepOps cluster to unlock its full potential and fuel groundbreaking discoveries.

    \section{Specialized Configurations}
    \subsection{GPUs with CUDA Support}
        While GPUs provide the raw horsepower for your DeepOps Slurm cluster, it's the carefully optimized libraries that truly turn this power into deep learning magic. These libraries act as your tools, offering pre-built functions and routines specifically designed to leverage the capabilities of GPUs, significantly accelerating your workflows.
        \subsubsection{CUDA: The Key to GPU Parallelism}
            CUDA (Compute Unified Device Architecture) is a parallel computing platform and programming model created by NVIDIA. It empowers developers to harness the massively parallel processing capabilities of GPUs, enabling them to execute thousands of computations simultaneously.
            CUDA-enabled GPUs offer significant performance gains for deep learning tasks, which often involve large-scale matrix operations and massive datasets.
        \subsubsection{Accelerating Deep Learning Libraries:}
            CUDA-accelerated libraries are specifically designed to leverage GPUs for deep learning computations, further boosting performance:
            \begin{itemize}
                \item cuDNN (CUDA Deep Neural Network library): Imagine having a toolbox specifically designed for building advanced structures. That's what cuDNN offers. This CUDA Deep Neural Network library provides highly optimized primitives for the fundamental operations used in deep learning architectures. Think convolutions, pooling, activation functions, and matrix multiplications – all performed with lightning speed due to cuDNN's tight integration with NVIDIA GPUs. This translates to faster training times and smoother performance for your deep learning models.
                \item TensorRT: Once you've trained your powerful deep learning model, you want to unleash its potential in real-world applications. That's where TensorRT comes in. This optimization tool takes your trained model and shrinks its size while preserving its accuracy. It accomplishes this by converting the model into a format specifically designed for high-performance inference on various platforms, including GPUs and embedded devices. Think of it as creating a leaner, meaner version of your model, ready to tackle real-world tasks with blazing speed and efficiency.
                \item Beyond cuDNN and TensorRT: The CUDA ecosystem offers a wider range of libraries that can accelerate your deep learning workflow:
                \begin{itemize}
                    \item cuBLAS: Optimized linear algebra library for matrix operations.
                    \item cuFFT: Fast Fourier Transform library for signal processing and image analysis.
                    \item cuSPARSE: Sparse matrix library for solving problems with sparse data structures.
                    \item RAPIDS: RAPIDS is an open-source suite of GPU-accelerated data science and AI libraries with APIs that match the most popular open-source data tools. It accelerates performance by orders of magnitude at scale across data pipelines. It offers cuGraph, 
                \end{itemize}
            \end{itemize}
        
        \subsubsection{Choosing CUDA-Enabled GPUs}
            DeepOps clusters prioritize GPUs with robust CUDA support, typically from the NVIDIA Tesla or NVIDIA A100 series.
            These GPUs offer:
            \begin{itemize}
                \item High-performance CUDA cores for parallel processing.
                \item Large memory capacities to accommodate massive datasets and model parameters.
                \item Advanced features like Tensor Cores for accelerated matrix operations.
            \end{itemize}
        
        CUDA's role in DeepOps is pivotal, ensuring that researchers and developers can push the boundaries of deep learning with unprecedented speed and performance.

    \subsection{Cluster Management Software}
        NVIDIA DeepOps or similar frameworks facilitate cluster deployment, management, and optimization specifically tailored for deep learning workloads. Red Hat cluster suite, Kubernets, TrinityX are example of some high availability open source cluster management software. Rocks Cluster Distribution, Warewulf are some open source non-high availability cluster management software. Only some of these cluster management software are designed for deep learning related tasks. There are several commercial solutions also available such as Bright Cluster Manager, NVIDIA DGX, ClusterVisor.

    \subsection{Scheduling and Resource Management}
        High-performance computing (HPC) environments require specialized software to manage and schedule complex workloads across multiple nodes. HPC scheduler systems act as the conductors of this intricate dance, ensuring efficient resource allocation, optimized job execution, and overall system performance.
        \begin{itemize}
            \item Open-Source Schedulers:
            \begin{itemize}
                \item Slurm: A widely popular, mature, and feature-rich scheduler known for its scalability, fairness policies, and extensive customization options.
                \item Torque/Maui: Another popular option offering good performance and integration with various resource management systems.
                \item PBS Pro: A widely used scheduler with a strong focus on security and user management.
                \item OpenLava: A newer, open-source scheduler designed with cloud-native principles, offering flexibility and scalability for modern HPC environments.
            \end{itemize}
            \item Commercial Schedulers
            \begin{itemize}
                \item Platform LSF: A comprehensive suite for workload management and job scheduling, offering advanced features like workload optimization and resource prediction.
                \item Altair PBS Professional: A commercially supported version of PBS Pro with additional features like workload visualization and analytics.
                \item Bright Cluster Manager: A highly scalable scheduler with a focus on ease of use and integration with various HPC environments.
                \item Grid Engine (SGE): A mature and widely used scheduler offering good performance and support for diverse workloads.
            \end{itemize}
        \end{itemize}
        Slurm, a highly efficient job scheduler, allocates resources (CPUs, GPUs, memory) to different jobs in the cluster, ensuring optimal utilization.
    \subsection{Multi-GPU Systems}
        The heart of a deep learning cluster lies not only in its network of connected nodes, but also in the individual nodes themselves. This is where multi-GPU systems come into play, serving as the multi-engine locomotives that propel your deep learning workloads forward at breakneck speed. Multi-GPU systems offer faster training times by reducing communication overhead over the network across multiple nodes.

    \chapter{Installation and Configuration with NVIDIA DeepOps}
    NVIDIA DeepOps provides a robust foundation for your deep learning endeavors, but the true magic lies in optimizing it for your specific needs. This note explores key areas where you can fine-tune your DeepOps cluster to unlock even faster training, smoother inference, and maximized resource utilization. NVIDIA DeepOps is a toolkit designed to facilitate the deployment and management of GPU-accelerated infrastructure for deep learning workloads.
    It streamlines the setup process for deep learning clusters, providing automation and configuration scripts optimized for NVIDIA hardware and software.

    \section{Prerequisites}
        \begin{itemize}
            \item Ensure the hardware components (GPUs, CPUs, memory, networking) are perfect.
            \item Install a supported Linux distribution (NVIDIA DGX OS 4, 5, Ubuntu 18.04 LTS, 20.04, 22.04 LTS, CentOS 7, 8) installed on each node. We have started with Ubuntu 22.04 LTS.
            \item Ensure Git is installed on the master node.
            \item Create a user named \texttt{ubuntu} in all nodes with same password.
            \item Note down hostname and IP addressess of all nodes.
            \item Make sure passwordless SSH acess exits across the nodes. To setup passwordless SSH access from master node to other nodes follow the below steps-
            \begin{enumerate}
                \item \textbf{Generate an SSH Key Pair}: 
                \begin{itemize}
                    \item Open a terminal on your local machine.
                    \item Type \textcolor{purple}{ssh-keygen} and press Enter.
                    \item Choose a file path to save the keys (default is usually fine).
                    \item Optionally, create a passphrase for added security (just pressing the Enter button is also fine).
                \end{itemize}
                \item \textbf{Copy the Public Key to the Remote Server}:
                \begin{itemize}
                    \item Use the ssh-copy-id command-
                    \begin{lstlisting}
                    ssh-copy-id ubuntu@remote_host
                    \end{lstlisting}
                    \item Enter your password when prompted.
                \end{itemize}
                \item \textbf{Verify Passwordless Login}:
                \begin{itemize}
                    \item Attempt to SSH into the server without a password:
                    \begin{lstlisting}
                    ssh ubuntu@remote_host
                    \end{lstlisting}
                    \item If successful, you'll log in directly without password prompts.
                \end{itemize}
            \end{enumerate}
            For a detailed guidance on this topic, we recommend you to visit this beautiful \href{https://www.digitalocean.com/community/tutorials/how-to-set-up-ssh-keys-on-ubuntu-20-04}{blog written by Digital Ocean}.
        \end{itemize}

    \section{Setup NVIDIA DeepOps}
        \subsubsection{On Master Node}
        \begin{enumerate}
            \item \textbf{Cloning DeepOps Repository}: Clone the NVIDIA DeepOps repository from the official GitHub repository to the head node or a designated management node by running the following command from terminal-
            \begin{lstlisting}
            git clone https://github.com/NVIDIA/deepops.git
            \end{lstlisting}
            \item \textbf{Set up your provisioning machine}: This will install Ansible and other software on the provisioning machine which will be used to deploy all other software to the cluster. For more information on Ansible and why we use it, consult the \href{https://github.com/NVIDIA/deepops/blob/master/docs/deepops/ansible.md}{Ansible Guide}. Go to the deepops directory and run the setup script.
    \begin{lstlisting}
    # Install software prerequisites and copy default configuration
    cd deepops
    ./scripts/setup.sh
    \end{lstlisting}
        \item \textbf{Create and edit the Ansible inventory}: Ansible uses an inventory which outlines the servers in your cluster. The setup script from the previous step will copy an example inventory configuration to the config directory. 
    \begin{lstlisting}
    # Edit inventory
    # Add Slurm controller/login host to `slurm-master` group
    # Add Slurm worker/compute hosts to the `slurm-node` groups
    nano config/inventory
    
    # (optional) Modify `config/group_vars/*.yml` to set configuration parameters
    \end{lstlisting}
            Add the hostnames and IP addresses of the nodes at appropriate positions of the file. See an example of inventory file here - \href{https:github.com/arindammajee/NVIDIA-DeepOps-Cluster/config/inventory}{https:github.com/arindammajee/NVIDIA-DeepOps-Cluster/config/inventory}
            Note: Multiple hosts can be added to the slurm-master group for high-availability. You must also set \textcolor{purple}{$slurm\_enable\_ha: true$}. in $config/group\_vars/slurm\_cluster.yml$. For more information about HA Slurm deployments, see: \href{https://slurm.schedmd.com/quickstart_admin.html#HA}{$https://slurm.schedmd.com/quickstart\_admin.html\#HA$}
            If you intend to configure \href{https://www.nvidia.com/en-us/technologies/multi-instance-gpu/}{Multi-Instance GPU}, consult the \href{https://github.com/NVIDIA/deepops/blob/master/docs/slurm-cluster/nvml.md}{Slurm NVML documentation}.
        \item To verify the configuration run the following command from deepops directory.
    \begin{lstlisting}
        ansible all -m raw -a "hostname"
    \end{lstlisting}
    And if everything is fine, you will see a result like Figure \ref{fig:deepops_verify_configuration}
    \begin{figure}
        \centering
        \includegraphics[scale=0.35]{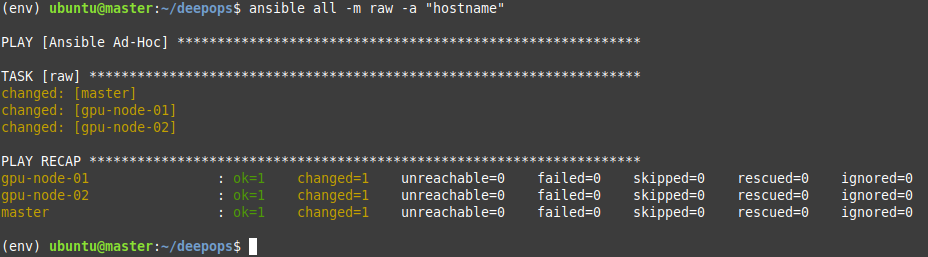}
        \caption{Caption}
        \label{fig:deepops_verify_configuration}
    \end{figure}

    \item Edit the \texttt{slurm-cluster playbooks/slurm-cluster.yml} file to configure your cluster. You can set \texttt{$install\_open\_ondemand$} to \texttt{True}. This will allow you to access your cluster remotely through a web-browser. But it is not recommended.
    \item Install Slurm by running below command and enter the sudo password of nodes.
\begin{lstlisting}
# NOTE: If SSH requires a password, add: `-k`
# NOTE: If sudo on remote machine requires a password, add: `-K`
# NOTE: If SSH user is different than current user, add: `-u ubuntu`
ansible-playbook -l slurm-cluster playbooks/slurm-cluster.yml -K
\end{lstlisting}
    \item After finishing the installation (it may take up to 2-4 hours depending on your internet speed), log in into the master node and run \texttt{sinfo} to check information about your cluster.
    \item You may try running \href{https://github.com/NVIDIA/deepops/tree/master/docs/slurm-cluster#slurm-validation}{SLURM-Validation test} to validate your cluster setup. To test all the nodes just run -
\begin{lstlisting}
ansible-playbook -l slurm-cluster playbooks/slurm-cluster/slurm-validation.yml
\end{lstlisting}
    \item Write an Ansible playbook to install your deep learning libraries across all the nodes. This will help you to install your libraries through a single command.
        \end{enumerate}
        Refer to \href{https://github.com/NVIDIA/deepops/tree/master/docs/slurm-cluster}{NIVIDIA DeepOps} for a detailed guide on setup a SLURM based DeepOps cluster. To setup a single node cluster you may refer to NVIDIA DeepOps's official blog \href{https://developer.nvidia.com/blog/deploying-rich-cluster-api-on-dgx-for-multi-user-sharing/}{Deploying Rich Cluster API on DGX for Multi-User Sharing}.

    \chapter{Accessing and Submitting Jobs}
    \section{Accessing the Cluster}
    \subsection{SSH Access}
        Users can access the cluster via Secure Shell (SSH) by using the command:
        \begin{lstlisting}
        ssh username@cluster_ip_address
        \end{lstlisting}
        Upon successful authentication, you will gain access to the cluster's command-line interface.
        Some clusters might offer a web-based interface or dashboard facilitating job submission, monitoring, and management.

    \section{Submitting Deep Learning Training Jobs}
    Running computationally demanding tasks on HPC clusters requires efficient resource management. This is where Slurm comes in, a powerful and open-source workload manager and a highly efficient job scheduler used on many HPC systems. Slurm helps users submit jobs, allocate resources like CPU cores and memory, and monitor their progress. Slurm offers a multitude of features beneficial for both users and administrators. Users can track job progress and resource utilization, submit parallel jobs across multiple nodes, and manage dependencies between tasks. Administrators can define different queues for prioritizing jobs, set resource limits, and implement advanced scheduling algorithms.
    
    By leveraging Slurm's capabilities, HPC users can maximize their productivity and optimize resource usage, leading to faster completion times and cost-effective computations.

    \subsection{Useful commands in SLURM}
    To get detailed idea of all SLURM commands, we request you to refer SLURM Documentation. Here, we will briefly discuss about only few commands which are most needed.
    \begin{itemize}
        \item \textbf{\textcolor{purple}{sinfo}}: View information about nodes and partitions (docs)
        \begin{table}[htbp]
            \centering
            \begin{tabularx}{\textwidth}{|>{\RaggedRight\arraybackslash}X|>{\RaggedRight\arraybackslash}X|}
            \hline
            \textbf{Option} & \textbf{Description} \\ \hline
            -o, --format=<options>	& Output format to display \\ \hline
            -l, --long	& Show more available information \\ \hline
            -N, --Node	& Show information in a node-oriented format \\ \hline
            -n, --nodes=<hostnames>	& Filter by host names (comma-separated list) \\ \hline
            -p, --$partition=<partition\_list$>	& Filter by partitions (comma-separated list) \\ \hline
            -t, --$states=<state\_list$>	& Filter by node states (comma-separated list) \\ \hline
            -s, --summarize	& Show summary information \\ \hline
            \end{tabularx}
            \caption{Options for sinfo command}
            \label{tab:sinfo}
        \end{table}

        \item \textbf{\textcolor{purple}{salloc}}: Obtain a job allocation for interactive use (docs)
        \begin{table}[htbp]
            \centering
            \begin{tabularx}{\textwidth}{|>{\RaggedRight\arraybackslash}X|>{\RaggedRight\arraybackslash}X|}
            \hline
            \textbf{Option} & \textbf{Description} \\ \hline
            -A, -\/-account=<account> &	Account to be charged for resources used \\ \hline
            -a, -\/-array=<index> &	Job array specification (sbatch only) \\ \hline
            -b, -\/-begin=<time> &	Initiate job after specified time \\ \hline
            -C, -\/-constraint=<features> &	Required node features \\ \hline
            -\/-cpu-bind=<type> &	Bind tasks to specific CPUs (srun only) \\ \hline
            -c, -\/-cpus-per-task=<count> &	Number of CPUs required per task \\ \hline
            -d, -\/-dependency=<state:jobid> &	Defer job until specified jobs reach specified state \\ \hline
            -m, -\/-distribution=<method[:method]> &	Specify distribution methods for remote processes \\ \hline
            -e, -\/-error=<filename> &	File in which to store job error messages (sbatch and srun only) \\ \hline
            -x, -\/-exclude=<name> &	Specify host names to exclude from job allocation \\ \hline
            -\/-exclusive &	Reserve all CPUs and GPUs on allocated nodes \\ \hline
            -\/-export=<name=value> &	Export specified environment variables (e.g., all, none) \\ \hline
            -\/-gpus-per-task=<list> &	Number of GPUs required per task \\ \hline
            -J, -\/-job-name=<name> &	Job name \\ \hline
            -l, -\/-label &	Prepend task ID to output (srun only) \\ \hline
            -\/-mail-type=<type> &	E-mail notification type (e.g., begin, end, fail, requeue, all) \\ \hline
            -\/-mail-user=<address> &	E-mail address \\ \hline
            -\/-mem=<size>[units] &	Memory required per allocated node (e.g., 16GB) \\ \hline
            -\/-mem-per-cpu=<size>[units] &	Memory required per allocated CPU (e.g., 2GB) \\ \hline
            -w, -\/-nodelist=<hostnames> &	Specify host names to include in job allocation \\ \hline
            -N, -\/-nodes=<count> &	Number of nodes required for the job \\ \hline
            -n, -\/-ntasks=<count> &	Number of tasks to be launched \\ \hline
            -\/-ntasks-per-node=<count> &	Number of tasks to be launched per node \\ \hline
            -o, -\/-output=<filename> &	File in which to store job output (sbatch and srun only) \\ \hline
            -p, -\/-partition=<names> &	Partition in which to run the job \\ \hline
            -t, -\/-time=<time> &	Limit for job run time \\ \hline
            \end{tabularx}
            \caption{Options for salloc, sbatch and srun command}
            \label{tab:salloc}
        \end{table}

        \item \textbf{\textcolor{purple}{sbatch}} - Submit a batch script for later execution (docs)
        \item \textbf{\textcolor{purple}{srun}} - Obtain a job allocation and run an application (docs)
        \item \textbf{\textcolor{purple}{scancel}} - Cancel a job. Syntax -
        \begin{lstlisting}
            scancel <job_id>
        \end{lstlisting}

        \item \textbf{\textcolor{purple}{squeue}} - View information about jobs in scheduling queue (docs)
        \begin{table}[htbp]
            \centering
            \begin{tabularx}{\textwidth}{|>{\RaggedRight\arraybackslash}X|>{\RaggedRight\arraybackslash}X|}
            \hline
            \textbf{Option} & \textbf{Description} \\ \hline
            -A, -\/-$account=<account\_list>$ &	Filter by accounts (comma-separated list) \\ \hline
            -o, -\/-format=<options> &	Output format to display \\ \hline
            -j, -\/-$jobs=<job\_id\_list>$ &	Filter by job IDs (comma-separated list) \\ \hline
            -l, -\/-long &	Show more available information \\ \hline
            -\/-me &	Filter by your own jobs \\ \hline
            -n, -\/-$name=<job\_name\_list>$ &	Filter by job names (comma-separated list) \\ \hline
            -p, -\/-$partition=<partition\_list>$ &	Filter by partitions (comma-separated list) \\ \hline
            -P, -\/-priority &	Sort jobs by priority \\ \hline
            -\/-start &	Show the expected start time and resources to be allocated for pending jobs \\ \hline
            -t, -\/-$states=<state\_list>$ &	Filter by states (comma-separated list) \\ \hline
            -u, -\/-$user=<user\_list>$ &	Filter by users (comma-separated list) \\ \hline
            \end{tabularx}
            \caption{Options for squeue command}
            \label{tab:squeue}
        \end{table}

        \item \textbf{\textcolor{purple}{scontrol}} - View and modify various aspects of SLURM's configuration and state.
        \begin{table}[htbp]
            \centering
            \begin{tabularx}{\textwidth}{|>{\RaggedRight\arraybackslash}X|>{\RaggedRight\arraybackslash}X|}
            \hline
            \textbf{Synatx} & \textbf{Usuage} \\ \hline
            scontrol show job $job\_id$ &	Display detailed information about a specific job. \\ \hline
            scontrol show nodes &	Displays details about the nodes in the SLURM cluster, such as their state, health, and available resources. \\ \hline
            scontrol update job $job\_id$ $attribute=value$ &	Filter by job IDs (comma-separated list) \\ \hline
            scontrol update nodename= <$node\_name$> state=<$desired\_state$> & Modify the state of nodes within the SLURM cluster. It allows administrators to change the state of compute nodes to make them available, offline, drained, or down for maintenance purposes. \\ \hline 

            \end{tabularx}
            \caption{Usuage of scontrol command}
            \label{tab:squeue}
        \end{table}

    \end{itemize}

    \subsection{Running an interactive job}
\begin{figure}
        \centering
        \includegraphics[scale=0.5]{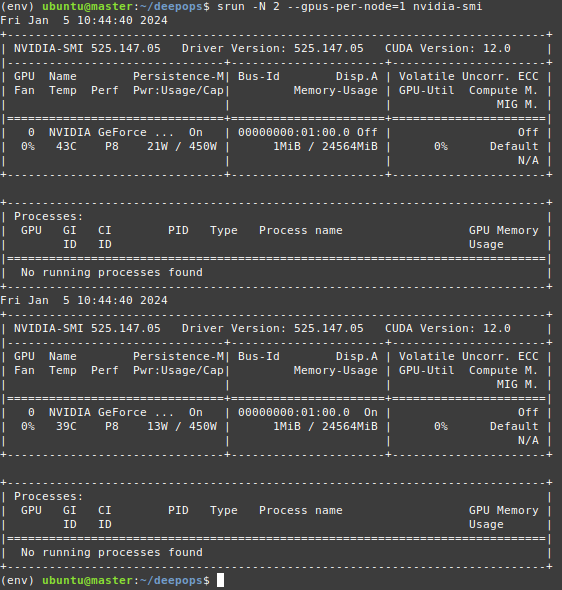}
        \caption{Output of running "srun -N 2 -\/-gpus-per-node=1 nvidia-smi"}
        \label{fig:deepops_srun_nvidia_smi}
    \end{figure}

    Loading the Slurm module gives you access to several commands, one of which is \textcolor{purple}{srun}. There are several different ways to use this command. To start off, we will begin an interactive job which asks for 1 compute node with 1 GPU for 1 hour.
    \begin{lstlisting}
    srun -N 1 --gpus=1 -t 01:00:00 --pty bash
    \end{lstlisting}
    To get 2 compute node with 1 GPU per node, we will below command-
    \begin{lstlisting}
    srun -N 2 --gpus-per-node=1 --pty bash
    \end{lstlisting}
    Please note that, the above command will request two nodes but, you will get access of only one node's terminal access. Instead of \textit{--pty bash }, if you use \textit{nvidia-smi}, you will get output from both nodes. For example, running
    \begin{lstlisting}
    srun -N 2 --gpus-per-node=1 nvidia-smi
    \end{lstlisting}
    gives us result as shown in \ref{fig:deepops_srun_nvidia_smi}
    
    \subsection{Job Submission using Slurm}
    Submitting a job involves writing a simple script specifying resource requirements, software environment, and the actual commands to execute. Slurm then schedules the job based on its priority, available resources, and queue policies. This ensures efficient utilization of the cluster by preventing resource conflicts and optimizing job scheduling. You can submit a job with the \textcolor{purple}{$sbatch$} command such as -
    \begin{lstlisting}
    sbatch job.slurm
    \end{lstlisting}

    \subsection{Job Submission Scripts for Deep Learning}
    Here's an example job submission script tailored for a deep learning task, highlighting GPU utilization and memory requirements:

    \begin{lstlisting}
    #!/bin/bash
    #SBATCH --job-name=deep_learning_job
    #SBATCH --partition=gpu
    #SBATCH --nodes=1
    #SBATCH --gres=gpu:1
    #SBATCH --cpus-per-task=8
    #SBATCH --mem=32G
    #SBATCH --time=24:00:00
    
   
    
    # Command to run deep learning training script
    python train.py --dataset /path/to/dataset --model resnet50 --batch-size 64 --epochs 50 --gpu
    
    # Additional commands for post-processing, saving models, etc.
    
        --job-name: Specifies the name of the job.
        --partition: Indicates the partition or queue for job execution (e.g., 'gpu' for nodes with GPUs).
        --nodes: Specifies the number of nodes required for the job.
        --gres=gpu:1: Requests one GPU for the job.
        --cpus-per-task: Sets the number of CPU cores per task.
        --mem: Allocates memory for the job (32GB in this example).
        --time: Specifies the maximum time the job is allowed to run (24 hours in this case).
    \end{lstlisting}

The script includes loading necessary modules (like CUDA) and activating a Python virtual environment before executing the actual deep learning training script (train.py). Adjust the parameters and commands according to your specific deep learning application.

\begin{lstlisting}
    # Load necessary modules (e.g., CUDA, Python virtual environment)
    module load cuda/XX.X  # Replace XX.X with the CUDA version
    source activate my_env  # Activate Python virtual environment
\end{lstlisting}

\subsection{Considerations}
    \begin{itemize}
        \item Resource Allocation: Specify appropriate resource requests (e.g., GPUs, memory) matching the requirements of the deep learning workload.
        \item Optimizations: Customize job scripts to optimize GPU usage, memory utilization, and other parameters based on the specific deep learning model and dataset.
        \item Virtual Enviournment: Always run your code in a virtual environment. This will keep your resources isolated from others.
        \item Use Default User: It is suggested to create a folder in the home directory of the default user \texttt{ubuntu} and keep your all resources (data, codes) inside that folder. This will help to utilize the storage resource properly.
    \end{itemize}

Conclusion:
Accessing the cluster and submitting deep learning jobs involves utilizing SSH or a web interface and leveraging Slurm job submission. Job submission scripts tailored for deep learning tasks help specify resource requirements, allowing efficient execution of training tasks on the cluster.

\chapter{Monitoring, Management and Maintenance}
    The journey of deep learning doesn't stop at building your DeepOps Slurm cluster. Just like any complex system, it requires constant monitoring, management, and maintenance to ensure optimal performance, prevent issues, and maximize its lifespan. This section dives into the essential practices for keeping your cluster humming smoothly.

    \section{Monitoring Cluster Status and Resource Utilization}
    \begin{itemize}
        \item Cluster Monitoring Commands: Utilize commands like \texttt{sinfo}, \texttt{squeue}, and \texttt{sacct} provided by SLURM to check the cluster's status, job queues, and historical job records. 
        \item Monitoring Tool: As part of the SLURM installation, Grafana and Prometheus are both deployed. These two tools are specially designed for monitoring HPC custer. The services can be reached from the following addresses:
        \begin{lstlisting}
            Grafana: http://<slurm-master-ip>:3000
            Prometheus: http://<slurm-master-ip>:9090
        \end{lstlisting}
        Visit respective services help pages for more details on them.
        \item Resource Monitoring Commands:
        Use system monitoring commands (top, htop, nvidia-smi) to monitor CPU, memory, and GPU utilization on individual nodes.
        GPU-specific commands like nvidia-smi provide details on GPU memory usage, temperature, and processes running on GPUs.
    \end{itemize}

    \section{Software Updates}
    Stay updated with the latest releases of software components crucial for deep learning tasks, including CUDA, cuDNN, deep learning frameworks (TensorFlow, PyTorch). Plan and execute software updates with minimal disruption to ongoing tasks by scheduling updates during low workload periods. It is suggested to use Ansible playbook to update packages across all the nodes.
    \section{Scaling Considerations}
    Please keep in mind that you can add or remove a node at any point of time. To add a node you have to just edit the inventory file and add install the DeepOps cluster again.

\chapter{Parallelism \& Optimizaion in Deep Learning}
    Deep learning models are becoming increasingly complex, demanding ever-greater computational resources to train effectively. This is where parallelism comes in, playing a crucial role in accelerating the training process and pushing the boundaries of what's possible. But with various types of parallelism available, choosing the right approach can be daunting. Let's delve into the different options and understand their strengths and limitations. Large deep learning models require significant amount of memory to train. Models require memory to store intermediate activations, weights etc.. In the modern machine learning the various approaches to parallelism are used to:
    \begin{itemize}
        \item Fit very large models onto limited hardware.
        \item Significantly speed up training.
    \end{itemize}
         
    We will first discuss about various types of parallelism techniques and their pros and cons and then look at how they can be combined to enable an even faster training and to support even bigger models. Various other powerful alternative approaches will be presented throughout this chapter.
    
    While the main concepts most likely will apply to any other framework, this chapter is focused on PyTorch-based implementations.

    \section{Parallelism}

    The following is the brief description of the main concepts that will be described later in depth in this document.
    \begin{itemize}
        \item \textbf{DataParallel (DP)} - the same setup is replicated multiple times, and each being fed a slice of the data. The processing is done in parallel and all setups are synchronized at the end of each training step.
        \item \textbf{TensorParallel (TP)} - each tensor is split up into multiple chunks, so instead of having the whole tensor reside on a single gpu, each shard of the tensor resides on its designated gpu. During processing each shard gets processed separately and in parallel on different GPUs and the results are synced at the end of the step. This is what one may call horizontal parallelism, as the splitting happens on horizontal level.
        \item \textbf{PipelineParallel (PP)} - the model is split up vertically (layer-level) across multiple GPUs, so that only one or several layers of the model are places on a single gpu. Each gpu processes in parallel different stages of the pipeline and working on a small chunk of the batch.
    \end{itemize}

    To get a better understanding of these parallelism we recommend you to go through these two beautiful blog - 
    \begin{itemize}
        \item \href{https://insujang.github.io/2022-06-11/parallelism-in-distributed-deep-learning/#parameter-server}{Parallelism in Distributed Deep Learning} - Blog by  Insu Jang, Ph.D. Candidate in CSE at the University of Michigan
        \item \href{https://huggingface.co/docs/transformers/v4.15.0/en/parallelism}{Model Parallelism} - Blog by Hugging Face.
    \end{itemize}

    These parallelism techniques can be combined with each others to accelerate training. Before we discuss about different other training and optimization techniques for training large models we recommend you to go through this survey paper- \href{https://arxiv.org/abs/2110.14883}{Colossal-AI: A Unified Deep Learning System For Large-Scale Parallel Training}

    \section{Different Training and Optimization Techniques}
    In this section we will discuss about different training and optimization techniques for deep learning models.
    \begin{enumerate}
        \item Fully Sharded Data Parallelism (FSDP): FSDP is a type of data-parallel training, but unlike traditional data-parallel, which maintains a per-GPU copy of a model’s parameters, gradients and optimizer states, it shards all of these states across data-parallel workers and can optionally offload the sharded model parameters to CPUs. We recommend you to go through these resources for a detailed understanding on FSDP.
        \begin{itemize}
            \item \href{https://engineering.fb.com/2021/07/15/open-source/fsdp/}{Fully Sharded Data Parallel: faster AI training with fewer GPUs}
            \item \href{https://pytorch.org/blog/introducing-pytorch-fully-sharded-data-parallel-api/}{Introducing PyTorch Fully Sharded Data Parallel (FSDP) API}
        \end{itemize}

        \item ZeRO: Zero Redundancy Optimizer(ZeRO) is a technique that eliminates memory redundancies by partitioning the optimizer, gradient, and parameters rather than replicating them to utilize the whole available memory. We recommend you to go through these beautiful resources for a better understanding on ZeRO - 
        \begin{itemize}
            \item \href{https://arxiv.org/abs/1910.02054}{ZeRO: Memory Optimizations Toward Training Trillion Parameter Models} - Microsoft Paper
            \item \href{https://arxiv.org/abs/2101.06840}{ZeRO-Offload: Democratizing Billion-Scale Model Training} - Microsoft Paper
            \item \href{https://oracle-oci-ocas.medium.com/zero-redundancy-optimizers-a-method-for-training-machine-learning-models-with-billion-parameter-472e8f4e7a5b}{Zero Redundancy Optimizers: A Method for training Machine Learning Models with Billion Parameters} - Blog from Research, Oracle Health \& AI
            \item \href{https://www.microsoft.com/en-us/research/blog/deepspeed-extreme-scale-model-training-for-everyone/}{DeepSpeed: Extreme-scale model training for everyone} - Blog from Microsoft Research
        \end{itemize}
    \end{enumerate}

    \section{Popular Frameworks for Large scale Training}
    There are several popular libraries for large scale model training. Some of these are -
    \begin{itemize}
        \item \href{https://huggingface.co/docs/transformers/en/accelerate}{Accelerate} - Developed by Hugging face
        \item \href{https://github.com/microsoft/DeepSpeed}{DeepSpeed} - Developed by Microsoft
        \item \href{https://github.com/facebookresearch/fairscale}{FairScale} -  by Facebook AI research.
    \end{itemize}

\chapter{Additional Resources}
    \section{Contact Us}
    \begin{itemize}
        \item We strongly recommend you to join this Google group to stay update for our updated version of this document and related discussions - \href{https://groups.google.com/g/tcg-crest-deep-learning/about}{https://groups.google.com/g/tcg-crest-deep-learning/about}
        \item For any kind of query, feel free to reach out to Mr. Arindam Majee via \href{mailto:majeearindam06072002@gmail.com}{majeearindam06072002@gmail.com} or visit \href{https://arindammajee.github.io/}{https://arindammajee.github.io/} to find a preferred method of contacting.
    \end{itemize}
    
    \section{NVIDIA DeepOps Documentation}
    \begin{itemize}
        \item Official Documentation: https://github.com/NVIDIA/deepops
        \item \href{https://www.nvidia.com/en-us/on-demand/session/gtcsj20-s22086/}{Deploying a Scalable GPU-as-a-Service Platform and Building a Deep Learning Project in Under 80 Minutes - NVIDIA OnDemand Video Session}
    \end{itemize}

\end{document}